\documentclass[9pt,twocolumn,twoside]{osajnl}

\journal{ol}

\setboolean{shortarticle}{true} 

\usepackage[percent]{overpic}
\usepackage{bm, upgreek}

\newcommand{\mum}{$\upmu$m }
\newcommand{\mumN}{$\upmu$m}
\newcommand{\assN}{As$_2$S$_3$}
\newcommand{\ass}{As$_2$S$_3$ }

\newcommand{\asse}{As$_2$Se$_3$ }

\usepackage[overlay,absolute]{textpos}
\newcommand\PlaceText[3]{%
	\begin{textblock*}{10in}(#1,#2)
		#3
	\end{textblock*}
}%

\title{Generation of 70~fs pulses at 2.86~$\bm{\upmu}$m from a mid-infrared fiber laser}

\author[1,*]{R. I. Woodward}
\author[1]{D. D. Hudson}
\author[1]{A. Fuerbach}
\author[1]{S. D. Jackson}

\affil[1]{MQ Photonics Research Centre, Macquarie University, New South Wales, Australia}

\affil[*]{Corresponding author: robert.woodward@mq.edu.au}

\ociscodes{(320.7090) Ultrafast lasers; (320.5520) Pulse compression; (190.4370) Nonlinear optics fibers.}

\begin{abstract}
We propose and demonstrate a simple route to few-optical-cycle pulse generation from a mid-infrared fiber laser through nonlinear compression of pulses from a holmium-doped fiber oscillator using a short length of chalcogenide fiber and a grating pair.
Pulses from the oscillator with 265~fs duration at 2.86~$\bm{\upmu}$m are spectrally broadened through self-phase modulation in step-index As$_2$S$_3$ fiber to 140~nm bandwidth, and then re-compressed to 70~fs (7.3 optical cycles).
These are the shortest pulses from a mid-infrared fiber system to date, and we note that our system is compact, robust and uses only commercially available components.
The scalability of this approach is also discussed, supported by numerical modeling.
\end{abstract}

\setboolean{displaycopyright}{false}

\begin{document}

\maketitle

\PlaceText{25mm}{9mm}{Vol. 42, Issue 23, pp. 4893-4896 (2017); https://doi.org/10.1364/OL.42.004893}

The generation of laser pulses comprising only a few cycles of the electric and magnetic fields creates substantial scientific and technological opportunities.
For example, such ultrashort pulses are enabling new time-resolved studies of atomic and molecular processes on unprecedented timescales and driving the development of tabletop extreme UV and attosecond pulse sources	through high-harmonic generation (HHG)~~\cite{Brabec2000}.
After significant progress in the near-infrared region, there is currently strong demand to push the wavelength of few-optical-cycle pulse sources to beyond 2~\mumN, into the mid-infrared (mid-IR): these wavelengths correspond to strong absorption resonances in many important organic materials, enabling further investigations and processing of new materials, and also offering advantages for HHG where the highest possible generated photon energy scales with the driving laser wavelength~\cite{Brabec2000}.

At present, a widely used approach to mid-IR few-cycle pulse generation is parametric wavelength conversion (e.g. optical parametric chirp pulse amplification) of ultrafast near-IR sources, often including a subsequent compression stage~\cite{Andriukaitis2011,Pupeza2015, Elu2017,Shumakova2016,Grafenstein2017}.
This route has yielded mid-IR few-cycle lasers with remarkable performance, but their significant complexity and cost limit widespread practical applications.
An alternative simpler approach is the direct development of mid-IR ultrafast oscillators.
Bulk transition-metal doped II-VI semiconductors such as chromium- and iron-doped sulfide and selenide offer direct access to the 2--6~\mum region, but while 46~fs pulses have been reported from mode-locked Cr:ZnS systems, such pulse generation has only been demonstrated thus far up to $\sim$2.4~\mumN~\cite{Mirov2015}.

One highly promising route is the recent emergence of ultrafast rare-earth-doped fluoride fiber lasers, which bring the benefits of fiber laser technology (compact setups, simple thermal management, high beam quality etc.) to the mid-IR region.
Mode-locked holmium- and erbium-based fiber lasers have been demonstrated at 2.7--2.9~\mum wavelengths~\cite{Hu2015b,Duval2015,Antipov2016a,Wei2017b,Zhu2017a} (including subsequent extension to 3.6~\mum using Raman soliton self-frequency shift~\cite{Duval2016}), producing tens of nanojoule energy pulses as short as 160~fs.
Due to the limited gain bandwidths of holmium and erbium ions (which set the minimum mode-locked pulse width via the transform limit), however, to achieve few-cycle pulses from mid-IR fiber lasers, additional extra-cavity nonlinear compression is required.

Nonlinear compression is a widely used technique involving nonlinear pulse propagation (e.g. in a gas-filled capillary or optical fiber) to broaden the spectral bandwidth by self-phase modulation (SPM), with linearizion of the chirp due to dispersion, and finally linear dispersion compensation (e.g. using diffraction gratings, a prism pair or chirped mirrors) to reduce the pulse width~\cite{Tomlinson1984}.
For example, at near-IR wavelengths, continuum generation in an all-normal dispersion silica photonic crystal fiber and chirped mirror compression recently resulted in sub-two-cycle-pulses~\cite{Heidt2011}.

In this Letter, we propose and experimentally demonstrate a simple route to few-cycle pulse generation from mid-IR fiber lasers using nonlinear compression in a short length of highly nonlinear step-index chalcogenide fiber.
This results in pulses as short as 70~fs (7.3 optical cycles) at 2.86~\mumN, which to our knowledge are the shortest pulses to date from a mid-IR fiber system by more than a factor of two.

\begin{figure}[bt]
	\centering
	\includegraphics{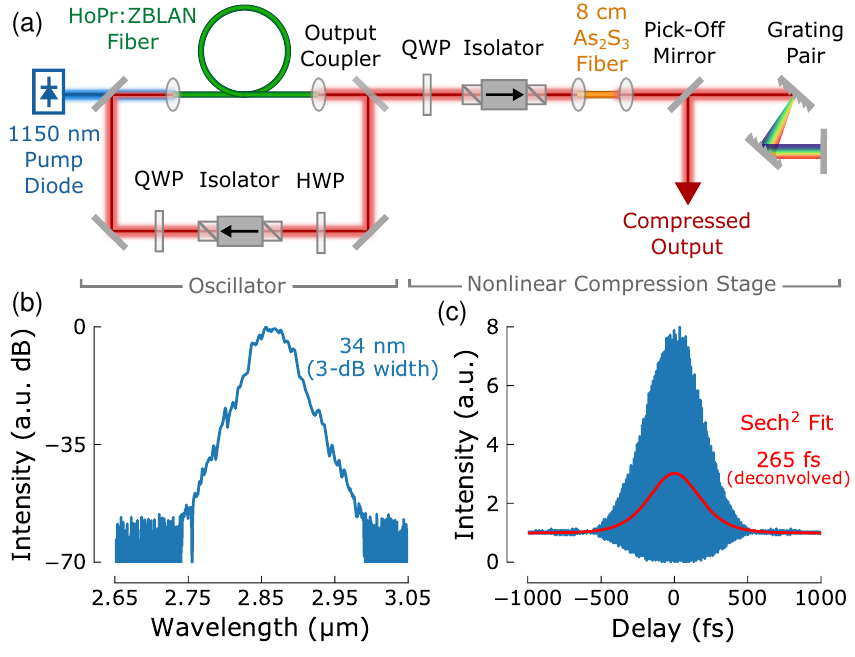}
	\caption{(a) Experimental setup. Mode-locked oscillator properties: (b) optical spectrum and (c) autocorrelation trace.}
	\label{fig:setup}
\end{figure}

A schematic of our laser system is shown in Fig.~\ref{fig:setup}(a).
The oscillator is a ring cavity including 3~m double-clad holmium-doped (3.5~mol\%) ZBLAN fiber (13~\mum core diameter, 0.13 core NA, 125~\mum octagonal cladding), pumped by an 1150~nm diode. 
Gain at 2.86~\mum arises from the Ho $^5I_6\rightarrow^5I_7$ transition, which is self-terminating due to the lower laser level possessing a longer lifetime than the upper state.
To overcome this limitation, the holmium fiber is co-doped with 0.25~mol\% praseodymium to depopulate the $^5I_7$ level in holmium due to highly resonant energy transfer to the closely spaced praseodymium $^3F_2$ level~\cite{Antipov2016a}.
The fiber tips are angle-cleaved to suppress parasitic feedback and a dichroic mirror with 43\% transmission is used as the output coupler.
Inclusion of a quarter and half waveplate (QWP \& HWP) in the cavity, in addition to the input polarizer of the isolator, creates an artificial saturable absorber through the well-known nonlinear polarization rotation (NPR) effect.
By careful adjustment of the waveplates, the cavity is biased to preferentially operate in a mode-locked state, suppressing CW emission.
Once the correct waveplate angles are identified (which in future will be algorithmically automated~\cite{Woodward_scirep_2016,Brunton2014}), mode-locking is reliably self-starting at $\sim$3~W pump power.

A stable pulse train is generated with 54~MHz repetition rate and up to 200~mW average power (3.7~nJ pulse energy).
The system operates at 2.865~\mum with 34~nm 3-dB bandwidth [Fig.~\ref{fig:setup}(b)], where the Ho:ZBLAN fiber is anomalously dispersive ($\beta_2\sim-109$~ps$^2$~km$^{-1}$) with nonlinear parameter $\gamma\sim0.2$~W$^{-1}$~km$^{-1}$.
Therefore, the laser operates in the soliton mode-locking regime and weak sidebands are observed in the output spectrum.
A custom-built interferometric two-photon absorption (TPA) based autocorrelator is used to measure the pulses, which are well fitted by a sech$^2$ shape (using least-squares fitting) with 410~fs width, corresponding to 265~fs pulse duration after accounting for the 0.647 autocorrelation deconvolution factor [Fig.~\ref{fig:setup}(c)].
The pulse time-bandwidth product (TBP) of 0.33 indicates the laser produces almost transform-limited pulses (i.e. very close to the fundamental limit of 0.315 for a sech pulse).

After the oscillator, a quarter waveplate is used to correct the small ellipticity of the output polarization and an isolator is installed to prevent destabilizing back reflections.
For the nonlinear compression stage, we require a medium for nonlinear pulse propagation while maintaining the coherence of the pulses. 
To maintain the simplicity benefits of fiber laser technology, we choose to use a length of step-index \ass chalcogenide fiber.
\ass exhibits more than two orders of magnitude greater nonlinearity than silica and fluoride glass and is strongly normally dispersive at 2.86~\mumN, which has been shown to promote coherent continuum generation compared to pumping at anomalously dispersive wavelengths~\cite{Heidt2011}.
We note that while octave-spanning supercontinua have recently been reported from \assN-based microwires and tapers~\cite{Hudson2011}, our aim here is to achieve broadening over a narrower range corresponding to the inverse bandwidth of our target pulse duration.
Step-index chalcogenide fiber is ideal for this purpose and we use off-the-shelf 5~\mum core diameter (0.28 NA) \ass fiber (IRFlex). 

Numerical modeling is used to guide the compressor design.
Briefly, pulse propagation is modeled using a generalized nonlinear Schr\"{o}dinger equation (GNLSE) that includes Raman scattering and optical shock, formulated in the frequency domain to include wavelength-dependent fiber properties.
Fiber properties including propagation constants $\beta(\lambda)$ (implicitly including group velocity dispersion and higher-order dispersion terms) and effective mode areas $A_\mathrm{eff}(\lambda)$ are computed for the fundamental HE$_{11}$ mode of a cylindrical step-index fiber using a vector eigenmode analysis~\cite{Snyder1983}, including the refractive index of \ass through a Sellmeier equation~\cite{Rodney1958}.
The nonlinear refractive index, $n_2$ of \ass at 2.86~\mum has not been reported, although at 1.55~\mum the value is commonly quoted in the range 3--6$\times10^{-18}$~m$^2$~W$^{-1}$ and a 2--3 times reduction is often suggested for $\sim$3~\mum operation ~\cite{Lamont2007} (e.g. 0.9$\times10^{-18}$~m$^2$~W$^{-1}$ at 3.1~\mum in Ref.~\cite{Hudson2014a}).
Based on this scaling, we propose $n_2=1.2\times10^{-18}$~m$^2$~W$^{-1}$ at 2.86~\mum for \assN, which is later verified by strong agreement between numerical and experimental results.

\begin{figure}[bt]
	\centering
	\includegraphics{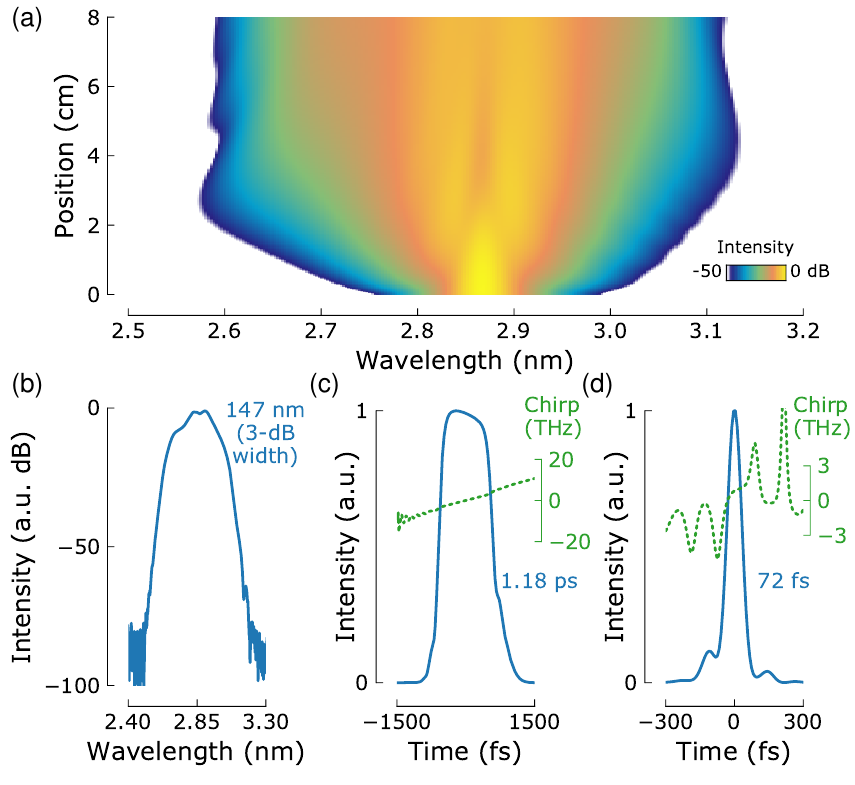}
	\caption{Numerically modeled nonlinear compression stage: (a) evolution of 3.5~kW, 265~fs sech$^2$ pulses in 8~cm \ass 5~\mum core diameter fiber; (b) spectrum and (c) pulse after fiber; (d) output pulse after double pass of an optimized grating pair compressor.}
	\label{fig:sim}
\end{figure}

At our laser operating wavelength, we compute for the 5~\mum core \ass fiber: group-velocity dispersion $\beta_2\sim294$~ps$^2$~km$^{-1}$, third-order dispersion $\beta_3\sim0.08$~ps$^3$~km$^{-1}$ and nonlinear parameter $\gamma\sim51$~W$^{-1}$~km$^{-1}$.
The simulated propagation of a 265~fs sech-shaped pulse (3.5~kW peak power) is shown in Fig.~\ref{fig:sim}(a).
Only a short length of a few centimeters is required to observe significant spectral broadening and we find that the SPM-induced nonlinear chirp is rapidly linearized due to the strong normal dispersion.
After 8~cm, the pulse 3-dB bandwidth is increased to 147~nm [Fig.~\ref{fig:sim}(b)] and the pulse dispersively broadens to 1.18~ps [Fig.~\ref{fig:sim}(c)].
Importantly, the chirp across the pulse is highly linear, which can be compensated using a linear compression technique.
Propagation over longer distances does not further broaden the spectrum, although does result in additional temporal broadening.
Therefore, it is preferable to use the shortest possible \ass fiber length for this nonlinear stage to minimize excessive pulse broadening and significant higher-order chirp accumulation from higher-order dispersion.

For dispersion compensation, we employ a double-pass diffraction grating pair with 70 lines per mm gratings.
The effect of the grating compressor is simulated by convolving the simulated pulses from the \ass fiber with the spectral phase function of a grating pair (after Ref.~\cite{Treacy1969}), where the grating pair separation is optimized to $\sim$5~cm to minimize the pulse duration.
The simulated pulses are thus compressed to 72~fs duration (TBP$=0.39$) and the output pulse shape is excellent, with only a very small pedestal [Fig.~\ref{fig:sim}(d)].
This slight pedestal and the deviation from the bandwidth-limited duration of 59~fs can be attributed to residual nonlinear / higher-order chirp that cannot be compensated using a simple grating pair.

\begin{figure}[bt]
	\centering
	\includegraphics{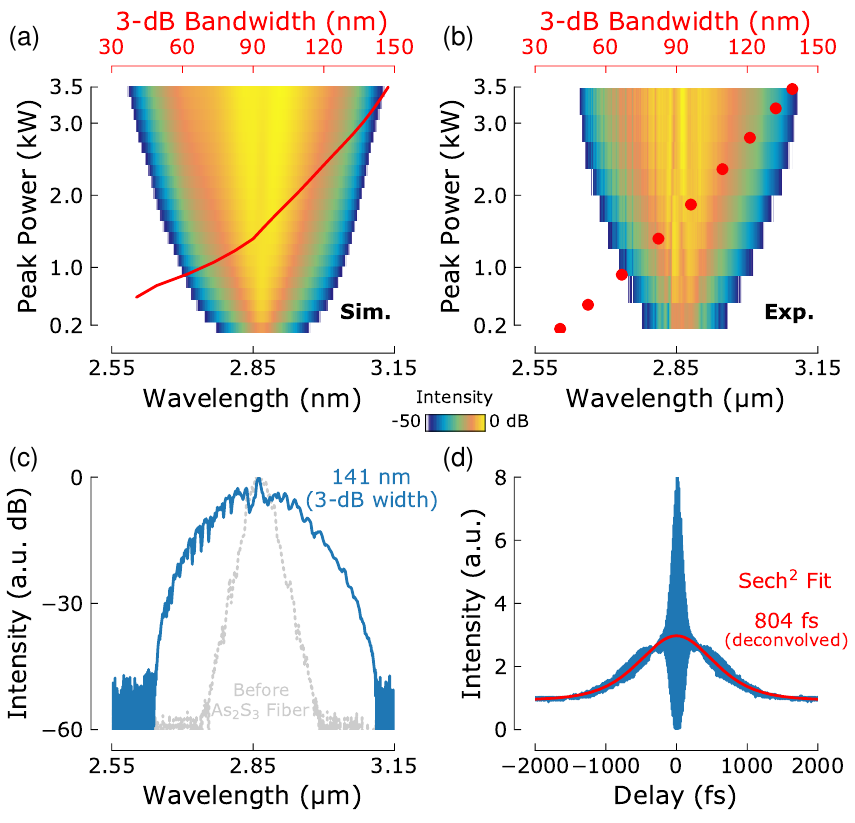}
	\caption{Spectral broadening as function of launched peak power: (a) numerical simulation; (b) experimental. Experimentally recorded (c) optical spectrum and (d) autocorrelation trace after \ass fiber for 3.5~kW launched power.}
	\label{fig:power_evol}
\end{figure}

Based on this design, we experimentally implement the nonlinear compression stage using an 8~cm length of \ass fiber.
The ends are planar cleaved using a diamond scribe and a 6~mm focal length zinc selenide aspheric lens is used to couple light into the core.
While the oscillator can deliver over 12~kW peak power, the maximum that can be coupled into the fiber is 3.5~kW, due to 40\% isolator loss, 20\% lens transmission loss, 17\% Fresnel reflection from the air-\ass interface ($n=2.417$ at 2.86~\mum) and 25\% mode-field mismatch loss.
With ongoing improvements to available mid-IR optical components, these losses could be minimized in future.

Spectral broadening after the \ass fiber is measured as a function of launched power and compared to the simulated results [Fig. \ref{fig:power_evol}(a)-(b)].
Due to the highly modulated nature of the measured spectra, widths are computed by first fitting a sech shape to the data.
Very good agreement is noted between the simulated and measured spectral widths, validating our model and the assumed nonlinear refractive index for \assN.
At the maximum launched power of 3.5~kW, the experimentally measured 3-dB spectral width is 141~nm [Fig.~\ref{fig:power_evol}(c)] and the broadened pulse is measured to be 804~fs [Fig.~\ref{fig:power_evol}(d)].
The interferometric autocorrelation trace only exhibits resolved fringes at the center, which is a classic signature of a chirped pulse.

\begin{figure}[bt]
	\centering
	\includegraphics{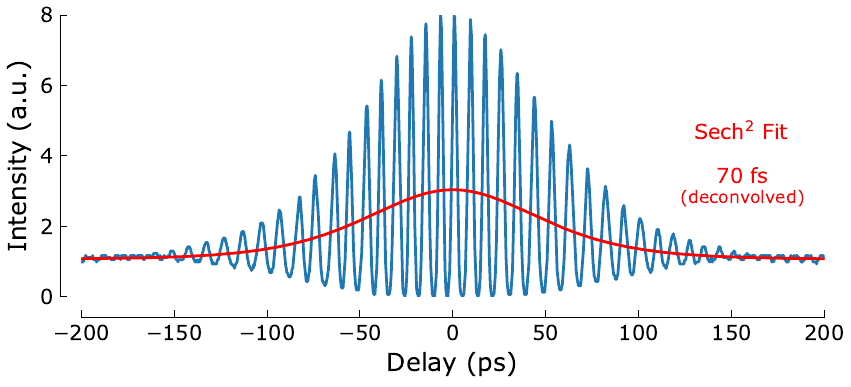}
	\caption{Experimental autocorrelation trace for compressed pulses after the grating pair.}
	\label{fig:compressed}
\end{figure}

The grating pair is assembled, operating at the blaze angle for maximum diffraction efficiency (measured to be 90\% per pass), with a preceding D-shaped mirror to pick-off the slightly offset reflected beam as the system output.
Grating compression is a linear effect and the reflected pulse spectrum in unchanged.
Temporally, however, we observe strong pulse compression as expected.
At an optimized separation of 5.5~cm, the output pulse duration is as short as 70~fs [Fig.~\ref{fig:compressed}], which corresponds to only 7.3 optical cycles (a single cycle at 2.865~\mum is 9.56~fs, as seen by the fringe spacing in the interferometric autocorrelation trace).
The clean autocorrelation trace exhibits no distortion, suggesting high pulse quality and we note that the compressed duration is very close the bandwidth-limited duration of 61~fs; the output pulse TBP is 0.36.
The experimental compression is in good agreement with the numerical model, confirming a reduction in the pulse width from the ultrafast fiber oscillator by a factor of four.

Pulse compression also results in a peak power enhancement to 6.3~kW (after loss through the \assN-air Fresnel reflection, fiber collimating lens, and 4 passes of diffraction gratings).
Improved efficiency could be achieved using a pair of Brewster-cut prisms for dispersion compensation, however, or even with an optimized length of anomalously dispersive ZBLAN fiber.

As an approach, we suggest that nonlinear compression in step-index chalcogenide fiber represents a promising route to practical few-cycle mid-IR fiber lasers.
It should be noted that chalcogenide fibers have previously been considered for pulse compression in the near-IR, although strong two-photon absorption across the semiconductor bandgap was revealed as a limiting loss mechanism~\cite{Fu2006}.
The situation is improved at mid-IR wavelengths as even two photons have insufficient energy to bridge the gap, suggesting greater opportunities for power scalability.
A recent study, however, identified the possibility of strong nonlinear absorption in \ass from two-photon absorption of valence electrons to the mid-gap Urbach extension, followed by linear absorption of excited states~\cite{Theberge2010}.
This is not likely to be a limiting factor for the short lengths required for nonlinear compression, although further work is needed to confirm this and to explore this loss mechanism.
For generating even shorter pulses, the mid-IR fiber oscillator could be amplified prior to compression as our numerical model suggests a quasi-linear increase in spectral bandwidth with increasing launch power, resulting in a shorter transform-limited duration.
Alternatively, \asse fiber could be considered which has a higher nonlinear refractive index than \ass~\cite{Fu2006}.

With continuing improvements to fiber splicing technology, chalcogenide fiber could be spliced directly to fluoride fibers, which would eliminate lens losses and reduce the Fresnel reflection.
This is a subject of ongoing research, in addition to the development all-fiber cavities with no free-space alignment sections, which is an important step towards the widespread availability of turn-key mid-IR fiber lasers.

In conclusion, we have proposed, numerically verified and experimentally demonstrated the generation of 70~fs pulses at 2.86~um (comprising only $\sim$7 optical cycles) from a fiber laser system, based on a mode-locked Ho:ZBLAN oscillator followed by nonlinear compression using a short length of \ass fiber and a grating pair.
This is the first demonstration of few-cycle mid-IR pulses from a fiber laser to our knowledge and we believe that nonlinear compression in chalcogenide step-index fiber is a simple and scalable route to enable even shorter pulses in the future.
Our diode-pumped laser system is compact and robust, bringing the benefits of few-cycle fiber laser technology to the mid-IR region and allowing for broader access to few-cycle pulses which will permit new investigations into light-matter interactions in the mid-infrared.

\section*{Funding Information}
MQRF Fellowship (RIW); Air Force Office of Scientific Research (FA2386-16-1-4030); Australian Research Council (DP170100531).

\bigskip


\end{document}